# Chemical bonding in chalcogenides: the concept of multi-centre hyperbonding

T. H. Lee* and S. R. Elliott*

Department of Chemistry, University of Cambridge, Lensfield Road, Cambridge, CB2 1EW, UK
E-mail: thl32@cam.ac.uk, sre1@cam.ac.uk

**Abstract**

The precise nature of chemical-bonding interactions in amorphous, and crystalline, chalcogenides is still unclear due to the complexity arising from the delocalization of bonding, and non-bonding, electrons. Although an increasing degree of electron delocalization for elements down a column of the periodic table is widely recognized, its influence on chemical-bonding interactions, and on consequent material properties, of chalcogenides has not previously been comprehensively understood from an atomistic point of view. Here, we provide a chemical-bonding framework for understanding the behaviour of chalcogenides (and, in principle, other lone-pair materials) by studying prototypical telluride non-volatile-memory, 'phase-change' materials (PCMs), and related chalcogenide compounds, via density-functional-theory, molecular-dynamics (DFT-MD) simulations. Identification of the presence of previously unconsidered multi-centre 'hyperbonding' (lone-pair–antibonding-orbital) interactions elucidates not only the origin of various material properties, and their contrast in magnitude between amorphous and crystalline phases, but also the very similar chemical-bonding nature between crystalline PCMs and one of the bonding subgroups (with the same bond length) found in amorphous PCMs, in marked contrast to existing viewpoints. The structure-property relationship established from this new bonding-interaction perspective will help in designing improved chalcogenide materials for diverse applications, based on a fundamental chemical-bonding point of view.

**Introduction**

Chalcogenide materials exhibit interesting properties for a wide range of applications, ranging from optical[1] or optoelectronic[2] applications, and encompassing applications of topological insulators,[3,4] to low-dimensional materials for electronics[5,6] or thermoelectric-generation devices.[7] This broad capability is feasible due to the wide tunability of material properties, depending on chalcogen-atom types, and by their generally wide compositional ranges for glass formation.[8]

Of all the chalcogenides, some tellurides, viz. 'phase-change' materials (PCMs), show unique material properties, i.e. ultrafast crystallization rates and large (opto-electronic) property contrasts between amorphous (*a*-) and crystalline (*c*-) phases.[9-14,15,16] The simultaneous occurrence of both these material properties is rather counterintuitive: the former is supposed to involve small structural differences between *a*- and *c*-phases, while the opposite trend would be expected for the latter. Consideration of electron delocalization is also seemingly necessary, in that a single Lewis structure, with bonding and non-bonding electron pairs, is unable fully to describe the electron distributions in both *a*- and *c*-PCM phases.[12,13] In order to answer this conundrum, chemical-bonding models, assuming a drastic change in the nature of chemical bonding between the two *a*- and *c*-phases, have been proposed,[14,17] and form the current mainstream point of view.

It is useful to classify existing models in terms of the length scale regarding electron delocalization. The resonant-bonding model,[12,14] based on a limited number of Lewis-conforming resonant structures, may be suitable for describing simple crystals with minimal disorder: naturally, structurally-disordered *a*-PCMs are beyond the validity of this model. Likewise, crystals with significant amounts of structural disorder are also difficult to incorporate within this model: PCM systems include metastable, vacancy-containing rocksalt-type $Ge_2Sb_2Te_5$ (*c*-GST)[18] (or, more generally, compositions along the pseudo-binary tie-line of GeTe–$Sb_2Te_3$[10]), Ag-In-Sb-Te (AIST),[19] I-V-VI$_2$-type compounds,[20] or *c*-GeTe with vacancies.[21] Although the relevant spatial scale is ill-defined, a measure of electron delocalization with substantial metallic character has been proposed, and termed 'metavalent bonding'.[22] A view based on atomic-orbital similarity has also been presented recently.[23] On the other hand, a molecular-orbital (MO) approach, involving notably three-centre, four-electron (3c/4e) interactions, or the valence-bond (VB) theory of hyperbonding, both focus on linear triatomic bonding geometries, and have been successfully adopted to describe hypervalent molecules in Chemistry.[24-26] Formation of linear triatomic bonding geometries has also been recognized in the solid state, in *a*-GST[13] and *c*-GeTe–$Sb_2Te_3$ alloys,[27] and a signature of electron delocalization within this bonding configuration is found in *a*-GST.[13] The advantage of this hyperbonding-interaction model over other models is that it can be characteristically linked to microscopic properties of the materials.

Here, we show that the concept of multi-centre, hyperbonding interactions completes the picture of chemical bonding in chalcogenides. Chemical-bonding interactions for *a*-/*c*-PCMs, and associated material properties, are comprehensively elucidated with this model. It is shown that, without invoking any other new type of bonding interaction, property contrasts between *a*- and *c*-phases of PCMs can be understood in terms of their naturally different extents of hyperbonding. This is in dramatic contrast with the existing theory for PCMs.

**Results and Discussion**

Figure 1 shows chemical-bonding-indicator data, calculated at the bond critical point (BCP), in simulated models of *a*-/*c*-GST. The trends for *a*-GST reveal that the charge density ($\rho_B$),[28] electron-localization function (ELF$_b$),[29] negative of the local energy density (LED),[30] or negative of the integrated crystal-orbital Hamilton population (ICOHP)[31] all increase monotonically with decreasing interatomic distance; this trend may be readily understood, given that the bonding is mainly covalent-like. This perspective is in accord with the observation that covalent-bonding interactions in molecules exhibit negative energy densities (while positive values are found for ionic or intermolecular interactions).[30] Other results, however, seemingly render this conclusion uncertain: for instance, charge densities at the BCP (~0.05 $e/a_0^3$) are rather small to be classified as being associated with purely covalent bonding;[32] and the values of $\nabla^2\rho_B$ are close to zero with a change of sign, an indication of the bonding character being intermediate between covalent (negative $\nabla^2\rho_B$) and ionic (positive $\nabla^2\rho_B$).[28] This presumably results from the additional contribution of metallicity, i.e. a tendency for delocalization of valence electrons.

Interestingly, a broad distribution of interatomic distances is also found for metastable rocksalt-type *c*-GST models (Figure 1), as for *a*-GST, due to structural disorder.[11,33] An important finding is that nearly all the bonding-character data-points for *c*-GST exactly overlap with parts of the wider range of data-points for *a*-GST (Figure 1a-e). Thus, in this overlap region, bonding characteristics are the *same* for *c*- and *a*-GST for the same bond length. This surprising similarity indicates that interatomic interactions in *c*-GST are indistinguishable from (some of) those in *a*-GST; more precisely, chemical-bonding interactions in *c*-GST belong to a subgroup of the broad spectrum of interactions existing in *a*-GST, but are *not* a different type of interaction to those present in *a*-GST. This finding is in strong contrast to the current consensus that the nature of bonding in *a*- and *c*-GST is inherently *dissimilar*,[11,14] which has been the basis for rationalizing the property contrasts in GST. Nevertheless, a meaningful difference exists between the peak positions (i.e. average bond length) of the interatomic-distance distributions: the peak positions for *c*-GST lie at longer interatomic distances compared to those for *a*-GST (Figure 1f). Such elongated interatomic distances are reminiscent of the relatively elongated axial bonds recently found for defective-octahedral configurations in *a*-GST.[13] The comparison in Figure 1f indeed reveals that the bond-length distribution for *c*-GST resembles the distribution for such axial bonds,[13] which is, along with the bonding-character data shown in Figure 1a-e, indicative of a similar bonding nature between bonds in *c*-GST and the axial bonds in *a*-GST.

In *a*-GST, the network structure is conventionally considered to consist of ordinary two-centre/two-electron (2c/2e) covalent bonds [Figure 2a(i)].[11,14] However, since Te lone pairs (LPs) can interact with neighbouring antibonding orbitals (via a hyperbonding interaction), LP delocalization-induced interactions with such antibonding orbitals should also be taken into account [Figure 2a(ii)]. The involvement of antibonding orbitals was presaged in ref. [13], where the COOP curves, shown in Figure 3e for axial-bonding Ge and Sb units in *a*-GST, clearly show the involvement of antibonding interactions for (lone-pair) states at the top of the valence band. According to a perturbative treatment for two-electron stabilization interactions,[34] the stabilization energy ($\Delta E_S$) of a LP ($\phi_A$) via the 3c/4e interaction is inversely proportional to the LP–antibonding energy difference ($\Delta E_{A-BC}$) (Figure 2b), while approximately being proportional to the extent of orbital overlap between the LP ($\phi_A$) and the interacting antibonding orbital ($\phi_{BC^*}$). For sufficiently strong 3c/4e interactions, pairs of (nearly) identical, (nearly) collinear bonds form (often denoted as 'hyperbonds' or 'ω bonds'[35]) [Figure 2a (iii)]. The schematic interaction-energy diagrams are depicted in Figure S7. Such hyperbond pairs correspond to the axial bonds in *a*-GST, and their bonding characteristics are clearly distinguishable from those of ordinary 2c/2e covalent bonds.[13] Distinct features include: i) a (near-) linear geometry of three bonded atoms, effectively maximizing the overlap between LPs and antibonding orbitals; ii) longer (and weaker) bonds; iii) higher polar covalency (i.e. ionicity), as visualized from the distorted shapes of maximally-localized Wannier functions (MLWFs), or the positions of MLWF centres being shifted towards Te atoms; and iv) stronger bonding-electron delocalization (Figure S6). These hyperbond characteristics observed in *a*-GST models coincide with those previously identified in molecules.[35] The term 'hyperbond' was coined[35] to emphasize the distinctive characteristics of the 3c/4e bonds relative to those of ordinary 2c/2e covalent bonds (see Methods for a detailed definition of hyperbonds).

Consequently, bonding interactions in *a*-GST can be of two different strong types, i.e. ordinary 2c/2e covalent bonds [B-C in Figure 2 (i)] and 3c/4e hyperbonds [A-B and B-C in Figure 2(iii)], along with a spectrum of weaker 3c/4e interactions [A⋯B in Figure 2 (ii)]. Due to the multi-centre nature of these interactions, atomic species can accommodate, with their four *s* and *p* atomic orbitals (AOs), up to six bonds and LPs in total, exceeding the ordinary Lewis maximum of four. This therefore addresses the question of how the coordination number in *a*-GST (and in *c*-GST) can approach six using their three *p* orbitals, corresponding to perfect octahedral coordination, and how the total number of bonds and LPs can reach five or six for hypervalent Ge or Sb atoms in *a*-GST.[13]

The proposed LP–antibonding interaction model (Figure 2b) may be validated as follows. First, bond lengths (or $ELF_b$ values) of 2c/2e bonds (B-C in Figure 2(i)) increase (or decrease) with the formation of 3c/4e interactions (Figure 2c), which conforms to the hyperbonding concept, as the antibonding level of the B-C bond becomes occupied via this interaction (Figure 2b). Second, the dependence of the hyperbonding tendency on the size of the band gap of materials (assumed to scale with $\Delta E_{A-BC}$) shows the expected trend (Figure 2d), i.e. a stronger hyperbonding tendency for materials with *smaller* band gaps, i.e. particularly tellurides (*a*-GST) compared to *a*-$Ge_2Sb_2S_5$ (*a*-GSS) and *a*-$Ge_2Sb_2Se_5$ (*a*-GSSe). Lastly, the involvement of Te LPs in forming hyperbonds is directly supported by the observation (Figure S5) that 96% (or 93%) of hyperbonds centred at Ge (or Sb) atoms have at least one, or two, three-fold coordinated Te atoms as ligands, each of which can donate the LP electrons required for the 3c/4e interactions (Figure 2a,b). These percentages are much higher than the overall percentage of three-fold Te atoms (~50%[13]) found in these models. It should be emphasized here that, although the hyperbonds and ordinary covalent bonds were defined when the corresponding $ELF_b$ value exceeds 0.5, other chemical-bonding indicators, or even a bond length, with a proper cutoff value can be equivalently used to define such bonds due to the definite relationships among those indicators (Figure 1a-e).

In *c*-GST, a significant difference with *a*-GST is that the crystalline symmetry requires Ge and Sb atoms to reside in (distorted) octahedral-ligand environments, allowing for the formation of a maximum of three perpendicular hyperbond pairs, i.e. six bonds. This means that the crystalline structure of metastable *c*-GST inherently provides conditions favourable for strong hyperbonding, with near-linear alignments of *p* orbitals. The sharp increase in the percentage of hyperbonds for *c*-GST, compared to *a*-GST (Figure 3a), is hence due to this crystal-structure amplification effect. The remaining percentage of non-hyperbonds corresponds to ordinary 2c/2e covalent bonds ((i) in Figure 2a) or weaker 3c/4e bonds ((ii) in Figure 2a). This finding, together with the high polar covalency of hyperbonds, is manifested in a higher degree of ionicity in *c*-GST than in *a*-GST, due to the charge transfer from Ge and Sb atoms to Te atoms (Figure S1), with the very pronounced increase of hyperbonds present in *c*-GST. This reveals an interesting character of hyperbonds that, although their electronegativity difference remains the same, simply switching between a σ (2c/2e) bond and a 3c/4e hyperbond changes the ionicity of the same bond (e.g. the B-C bond in Figure 2a).

It is now clear that the overall bonding difference between *a*- and *c*-GST arises from the relative abundances of the bonding types; 2c/2e covalent bonding predominates in *a*-GST, with a minor proportion of hyperbonds, while dominant multi-centre 3c/4e hyperbonding, with a minor proportion of 2c/2e covalent interactions, prevails for *c*-GST, amplified by its crystal symmetry. LP-delocalization-induced weak 3c/4e interactions [Figure 2a (ii)] constitute an essential component of the chemical-bonding interactions in GST, mediating between the two stronger interactions of 2c/2e bonds and of hyperbonds (Figure 2a). LP-delocalization-induced weak 3c/4e interactions could occur for any LP-containing material, but with different interaction strengths, depending on their electronic structures (as indicated in Figure 2b,d), and hence can have a more general importance.

We can link the hyperbonding concept to many material properties. The first to be considered is a salient feature of PCMs, namely the large optical-property contrast between *a*- and *c*-phases, e.g. of GST.[14] The relationship found here between hyperbonds and Born effective charges (BECs) is the key finding that associates these seemingly unrelated properties (Figure 3a,b). The BEC, averaged over all Ge, Sb, and Te atoms in *a*-GST (*c*-GST) models was found to be 2.72±0.98 (6.40±2.30), 3.72±1.76 (8.74±3.60), and -2.57±1.30 (-6.16±3.41), respectively. These values are comparable to those given in a previous report.[36] The averaged BEC values for the atomic constituents in *a*-GST are slightly higher than their nominal ionic charges, but the atomic constituents of *c*-GST exhibit much higher BEC values. The sharp increase of BEC values from *a*- to *c*-GST follows the similar trend of increasing ratios of proportion of hyperbonds to that of ordinary covalent bonds (Figure 3a), indicative of an intimate connection between BEC values and the presence of hyperbonds. To obtain a deeper insight into their relationship, we decomposed the whole distribution of BECs into contributions from atoms belonging to groups of atoms involved in forming a specific number of hyperbonds. As the AOs of each atomic constituent are involved in forming more hyperbond pairs, the atomic constituent tends to show a higher BEC value (Figure 3b). In other words, hyperbonding facilitates a larger BEC than does ordinary covalent bonding; the high sensitivity of electronic polarization to atomic displacements[37] therefore constitutes another characteristic of hyperbonding. Considering the reported proportional relation between BECs and dielectric constants,[38] it is the drastically increased number of hyperbonds in *c*-PCMs that escalates BEC and dielectric-constant values. Thus, crystal symmetry, permitting hyperbonding, is an essential requirement for large optical contrasts between *a*- and *c*-phases: the resulting linear atomic alignment hence plays a crucial role. The importance of aligned *p* orbitals in giving high dielectric constants was also discussed by Huang et al.,[39] supporting the present conclusion.

The formation, or decomposition, process of hyperbonds, depicted in Figure 2a, provides an additional insight into the dynamical properties of PCMs. The first example of interest is the collective bond-breaking behaviour in PCMs found from laser-assisted-evaporation experiments,[40] presented as supporting evidence for the existence of metavalent bonding.[22] The essential observation identified by us is that the reported multiple-event probability[40] (MEP) appears to scale with the hyperbond content. In view of the hyperbonding mechanism and the observations for *a*-GST, the bonding network of amorphous PCMs can be rather generally described by mixed bonding types, ranging from ordinary covalent bonding to strong hyperbonding (i.e. (i) to (iii) in Figure 2a). Given the capability of atom-probe tomography in differentiating those bonding types,[40] the observations in Ref. [40] of higher MEPs for higher Te contents in *a*-GeSe$_X$Te$_{1-X}$ (where *x* = 0.25, 0.5, 0.75) may be understood by the stronger hyperbonding tendency of Te atoms. On the other hand, the observation of much higher MEPs for any *a*-GeSe$_X$Te$_{1-X}$ samples than those for the *sp*$^3$-bonded (non-hyperbonding) *c*-InSb, and for the *p*-bonded (albeit, non-hyperbonding) *c*-GeSe, can be attributed to the presence, or absence, of multi-centre hyperbonding in each material, thereby enabling the differentiation between *a*- and *c*-PCMs, as well as between PCMs and non-PCMs, by considering the presence of hyperbonding in both *a*- and *c*-PCMs. Further associated discussion will be given in the following. In any case, the collective bond-breaking itself can be attributed to the unique response of strongly resonating hyperbonds.

Another finding is that, unlike in crystalline Ge-Sb-S (*c*-GSS) or Ge-Sb-Se (*c*-GSSe), interatomic distances in the telluride material, *c*-GST, show a continuous distribution without any significant gap (or dip) that, otherwise, separates contributions from strong and weak interatomic interactions (Figure 1f and Figure S2). This is because hyperbonding bridges the gap between strong 2c/2e covalent bonding and weak LP interactions. We therefore postulate that the presence of hyperbonding can provide low-energy-barrier routes for bond formation and breakage: for instance, the transition from an A: + B-C configuration ((i) in Figure 2a) to an A-B + :C configuration (as a result of the B-C bond breaking from (iii) in Figure 2a, leaving a LP on the C atom) leads to the formation of a new A-B bond at the expense of the B-C bond via an A-B-C hyperbond. This LP-delocalization-assisted formation of hyperbonds, and facile bond switching, is in line with the reported fast valence-charge redistribution and rapid crystallization in this material at high temperatures.[13,17] This behaviour is beneficial for fast SET operations for PC memory applications. Conversely, the presence of a fast bond-switching route explains why PCMs are, in general, poor glass formers, which is detrimental to the thermal stability of amorphous PCMs. The decisive role of tetrahedrally-bonded dopants in enhancing the thermal stability of amorphous PCMs[41] can be ascribed, from the hyperbonding perspective, to the suppression of these bond-switching processes by the dopants. Finally, a hint of a correlation with phonon anharmonicity is given by the recent observation by Lee et al.[42] that a strong phonon anharmonicity has its fundamental root in the cubic (rocksalt) crystalline structure, enabling a linear alignment of *p* AOs. The same condition for maximizing hyperbonding suggests that strong anharmonicity may therefore accompany pronounced hyperbonding. It is also interesting to note that strong phonon anharmonicity is often linked to the presence of shallow double-well potentials due to weak Peierls distortions,[33,43] which corresponds to the presence of (weak) hyperbonds.

Now let us compare with other chalcogenides containing different types of chalcogens, as already partly discussed for Figure 2d. On going from GSS to GSSe, and to GST, rather general trends in bonding characteristics are apparent: i) the difference between GSS and GSSe is relatively small, yet a drastic change occurs from GSSe to GST; and ii) the contrast between *a*- and *c*-phases becomes amplified. For instance, a rather small increase of the hyperbond ratio (Figure 3a) occurs from GSS to GSSe, followed by a much larger increment from GSSe to GST, together with the amplifying contrasts between *a*- and *c*-models. Although the contrast becomes weaker, basically the same trends are observed for the averaged BECs (Figure 3a). The origin of this distinction between chalcogenides may be described in terms of the factors affecting hyperbonding (Figure 2). The strength of hyperbonding interactions is determined by the energy separation between the energy levels of the interacting LP and antibonding orbitals, along with the extent of spatial overlap between these two orbitals; the lower the energy separation (and/or the larger the overlap of the orbitals), the stronger the interaction. Thus, the prevalence of hyperbonding in tellurides with their valence electrons being more delocalized, rather than in selenides or particularly sulphides, is therefore naturally understood. However, we found that the extent of *sp* hybridization of the constituent AOs *also* conveniently discriminates these differences. For the amorphous phases, the trend of *sp* hybridization, as measured by the distance from the central atom to its LP MLWF centre (WFC$_{LP}$), is clearly in inverse proportion to the degree of hyperbonding (Figure 3a). The same conclusion is reached for crystalline models (see Supplementary Information for details). This *sp* hybridization, stimulating stereochemical bonding activity of LPs,[44,45] leads to structural distortion, followed by the breakage of linear atomic alignments in *c*-GSS, and less so in *c*-GSSe, i.e. a loss of hyperbonding. The local atomic-coordination change involved in the distortion in the crystalline models is found to involve transitions from the initial octahedral coordination to lower coordinated polyhedral units, notably with trigonal-pyramidal geometry,[13] that can be described by the sequential decomposition processes of (iii) → (ii) → (i) in Figure 2a, along the three hyperbonding directions. This is not the case for *c*-GST (see Figure 3). The trend of stability of multi-centre hyperbonding interactions, along with the sustainability of the cubic crystalline form, is, therefore, in increasing order of GSS, GSSe, and GST. In fact, the *sp* hybridization of AOs itself is detrimental to the hyperbonding, as it diminishes the overlap between AOs compared to the case of three pure *p* orbitals being involved. The close correlation between the hybridization of AOs, multi-centre hyperbonding, and crystal structure therefore rationalizes why most PCMs are tellurides rather than sulphides or selenides. With the same reasoning, it can also be understood why sulphides, or selenides, of binary crystalline IV-VI compounds adopt the black-phosphorus crystalline structure, which is more distorted from its hypothetical cubic rocksalt structure than are tellurides with a rhombohedral structure (e.g. GeTe or SnTe).[44]

A distinction between chalcogenides is also found in their extent of electron delocalization (Figure 3d). Electron delocalization occurs to a significant extent in *a*-GST, best illustrated from the substantial increase of the spread ($\Omega$) of bonding MLWFs with the heavier Gp. VI elements (Figure 3d), an indication of increasing metallicity.[44] Such a metallic-like contribution explains the unusually small values of chemical-bonding indicators for *a*- and *c*-GST in Figure 1. The additional contribution of hyperbonds to delocalization (see Figure S6) is manifested in a broader standard deviation of $\Omega$ values for *a*-GST than for other amorphous chalcogenides. For *c*-GST, very significant electron delocalization, in comparison with that in *a*-GST, is expected because of a substantial degree of hyperbonding. The concept of multi-centre hyperbonding, therefore, explains the intermediate nature of interactions between covalent and metallic bonding in GST in terms of electron delocalization.

Another material system of interest is GeTe, another prototype PCM material. From the hyperbonding point of view (see Methods for more details), long bonds in *c*-GeTe may be described by a stabilization interaction between Te LPs in one Ge-Te layer and nearby Ge-Te antibonding states in another layer. However, the interaction is not strong (e.g. due to the lack of ideal overlap of relevant orbitals restricted by the crystalline structure), resulting in the formation of alternating one short- and one long-bond layers. A pair of short and long bonds with a linear triatomic bonding geometry can be regarded as an incomplete hyperbond pair, whose configuration lies between cases (ii) and (iii) in Figure 2a. As in GST, chemical-bonding indicators for *c*-GeTe also well overlap (some of) those for *a*-GeTe, regardless of the functionals used (Figure 4), indicating that chemical-bonding interactions in *c*-GeTe can also be described as a subgroup of bonding interactions in *a*-GeTe, as for GST.

We further considered two factors affecting the bonding in *c*-GeTe, namely pressure and the presence of atomic vacancies. Regardless of pressure or even the presence of cation vacancies (Figure 4 and see the Methods), the chemical-bonding indicators for both types of bonds in *c*-GeTe always remain within the range of values found in *a*-GeTe, as also found for GST. On the other hand, the presence of cation vacancies destroys the local symmetry around a Te LP, triggering stronger 3c/4e interactions, as revealed by the coupled shortening, and lengthening, respectively, of long, and short, bonds (Figure 4). As a result, stronger linkages across Ge-Te long-bond layers occur via the process of (ii) → (iii) in Figure 2a. Shortening of a small number of Ge-Te short bonds is also found. Although the overall changes in terms of atomic positions is marginal,[21] the detailed charge redistribution is significant, as indicated by the large disorder in the ELF distribution induced by the introduced vacancies. This disorder is induced as a way of minimizing the system energy following the formation of 3c/4e interactions. Therefore, multi-centre hyperbonding interactions provide a useful means of stabilizing defects in *c*-GeTe, and presumably also in GST. Such vacancy formation induces electron localization[50] near Te atoms, next to the vacancies, (Figure S8), and the positions of the volumes enclosed by high-ELF isosurfaces closely correspond to the positions of LPs predicted by the valence-shell electron-pair repulsion (VSEPR) theory for each local coordination of the Te atoms, an indication of the formation of stereochemically-active LPs accompanying vacancy formation.

The concept of hyperbonding described so far suggests an elementary three-body structural motif in chalcogenides, which consists of either two equal bonds (Figure 2a (iii)), or a pair of short and long bonds (Figure 2a (ii)), with a perfect, or close to, linear bonding geometry, respectively. The difference between the three configurations in Figure 2a (i.e. from (i) to (iii)) simply originates from their different strengths of hyperbonding interactions, whose variations are continuous rather than clearly differentiating between the configurations (Figure 2c). This often imposes a difficulty in, e.g., defining coordination numbers of atoms in relevant amorphous chalcogenides, and becomes the origin of severe overlaps between the first and second peaks in pair correlation functions.[13] Also, considering a lack of rigorous justification in disordered material systems, the notion of a 'Peierls distortion' for liquid, supercooled liquid, or glassy chalcogenides,[43] which has often been used to describe such linear bonding geometries, can be better rephrased as a 'weak hyperbonding interaction' (Figure 2b (ii)), and the presence, and material-dependent abundance, of such geometries in chalcogenide liquids or glasses can be rationalized by the theory of hyperbonding. Another interesting aspect of the hyperbonding model, as stated previously, is that the formation of both weak ((ii) in Figure 2a) and strong ((iii) in Figure 2a) hyperbonds results in the formation of energy levels near the top of the valence band with substantial *antibonding* character (Figure 2b). Accordingly, such antibonding-character states are observed for amorphous[13] and crystalline[51] chalcogenides where substantial linear triatomic motifs are present. The observation of an overall decrease in the antibonding character with vacancy formation, leading to a stabilization of crystalline cubic GST[51], can be naturally accounted for in the hyperbonding picture in terms of the reduced total number of hyperbonds with vacancy formation in the cubic crystalline structure.

The clear connections established between hyperbonding interactions, local atomic geometries, and material properties indicate that the structures of materials themselves may therefore provide guidelines for material selection for PCM, or thermoelectric, applications. As emphasised in the previous section, in order to achieve strong hyperbonding interactions in chalcogenides, material structures need to possess crystalline symmetries supporting linear triatomic bonding geometries. (Distorted) rocksalt or rhombohedral crystal structures, for example, satisfy this criterion, whose structures are commonly observed for most of the known phase-change, or thermoelectric, materials: they share a diversity of material properties characteristic of hyperbonding materials, such as high coordination numbers beyond the Lewis octet rule (i.e. the number of bond pairs plus LPs exceeds 4) and high Born-effective charges (or high dielectric constants), along with the aforementioned antibonding character.[10-16,22] The hyperbonding model hence rationalizes the presence of the preferred crystalline structures for PCM, and thermoelectric, applications. Apart from the distorted rocksalt *c*-GST and rhombohedral *c*-GeTe materials studied here, other relevant material systems may include hexagonal *c*-GST,[52] *c*-AIST,[19] I-V-VI$_2$-type compounds,[20] and crystalline chalcogenides with layered structures, e.g. rhombohedral V$_2$VI$_3$-type crystals of *c*-Sb$_2$Te$_3$, *c*-Bi$_2$Te$_3$, etc.[16] Layered-structure materials with similar compositions, yet whose structures lack such linear structural motifs (such as *c*-Sb$_2$Se$_3$), seldom exhibit these hyperbonding characteristics,[53] consistent with the hyperbonding perspective. In this respect, wider application of the hyperbonding concept for diverse crystals seems to be promising, and is worth future investigation. Narrow bandgaps (Figure 2b), and types of anions which induce minimal *sp* hybridization of cationic atomic orbitals,[44] are also important factors for such applications. Hence, $sp^3$-bonded glasses, or crystals, with tetrahedral-bonding geometries are representative of non-hyperbonding materials. The hyperbonding model, as outlined in Figure 2, is therefore seemingly able to account for most experimental/simulational observations reported so far, thereby providing a unifying framework for understanding broad classes of chalcogenides, while also being able to suggest useful guidelines for material selection for applications. In particular, the hyperbonding model appears able to establish a clear structure/property relationship, which is absent in other chemical-bonding models proposed so far.

**Conclusions**

The concept of multi-centre, lone-pair–antibonding hyperbonding interactions extends conventional chemical-bonding theory to give a correct understanding of the electronic structure and properties of chalcogenides. A combination of three-centre/four electron hyperbonding and ordinary two-centre/two-electron covalent bonding can elucidate the origin of chalcogen-dependent structural differences and material properties associated with heavy Group VI elements. The established connection between chemical bonding, crystal structure, and material properties provides a completely new perspective in understanding, and hence designing, chalcogenide materials for various applications, including phase-change-memory or thermoelectric-generation materials.

**Acknowledgements**

This work was supported by the UK Engineering and Physical Sciences Research Council (EPSRC) grants (EP/N022009, EP/M015130). The DFT calculations were performed using the ARCHER UK National Supercomputing service (HEC Materials Chemistry Consortium), which is funded by EPSRC (EP/L000202, EP/R029431), and the Cambridge High-Performance Computing Facility (Darwin).

**Author Information**

The authors declare no competing financial interests. Correspondence should be addressed to T. H. L. (thl32@cam.ac.uk) or S. R. E. (sre1@cam.ac.uk).

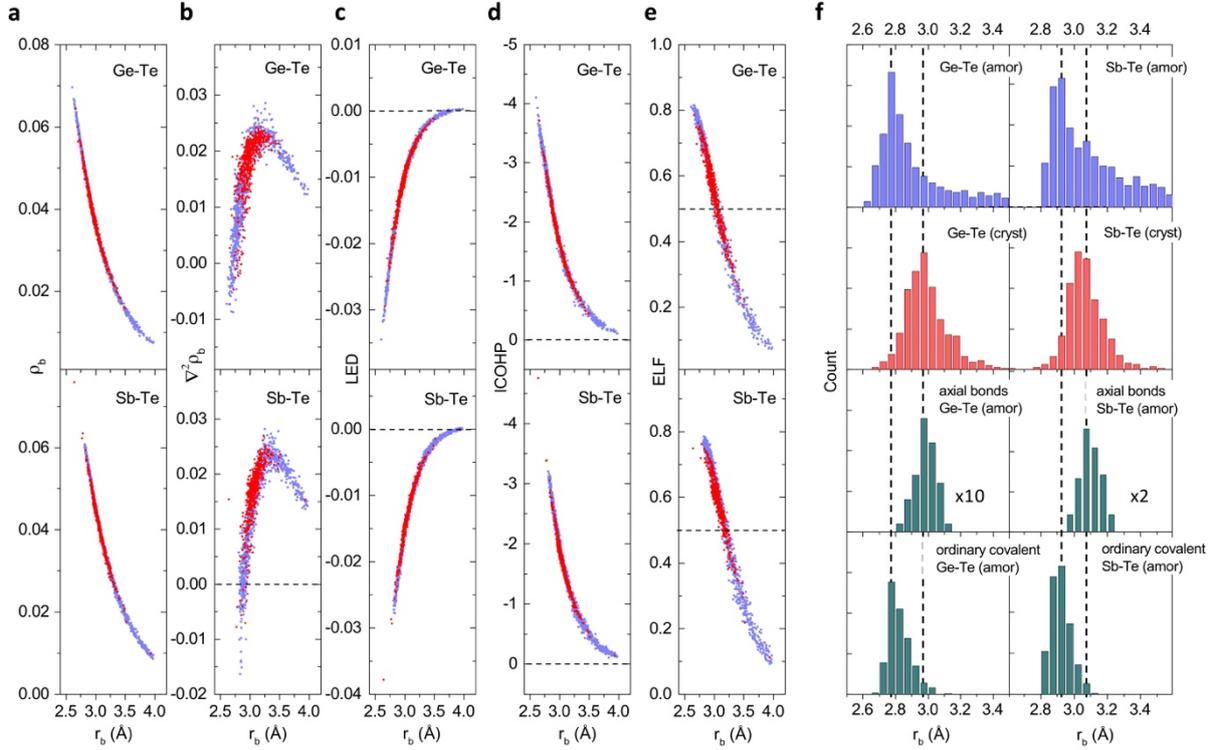

**Figure 1.** Chemical-bonding indicators and interatomic-distance distributions for amorphous and crystalline GST models. Data points (blue for amorphous, and red for crystalline, phases) evaluated at the bond-critical point for: (a) charge density ($e/a_0^3$); (b) Laplacian of the charge density ($e/a_0^5$); (c) local energy density (Hartree/$a_0^3$); and (e) ELF, where $e$ is the electron charge and $a_0$ is the Bohr radius. Data points for (d) ICOHP (eV) are also shown. (f) Distribution of interatomic distances for amorphous (top) and crystalline GST models (second from top). The axial bonds (second from bottom) and the remaining ordinary covalent bonds (bottom) in *a*-GST models show similar interatomic-distance distributions. Here, all models were generated, and characterized, with PBE exchange-correlation functionals.

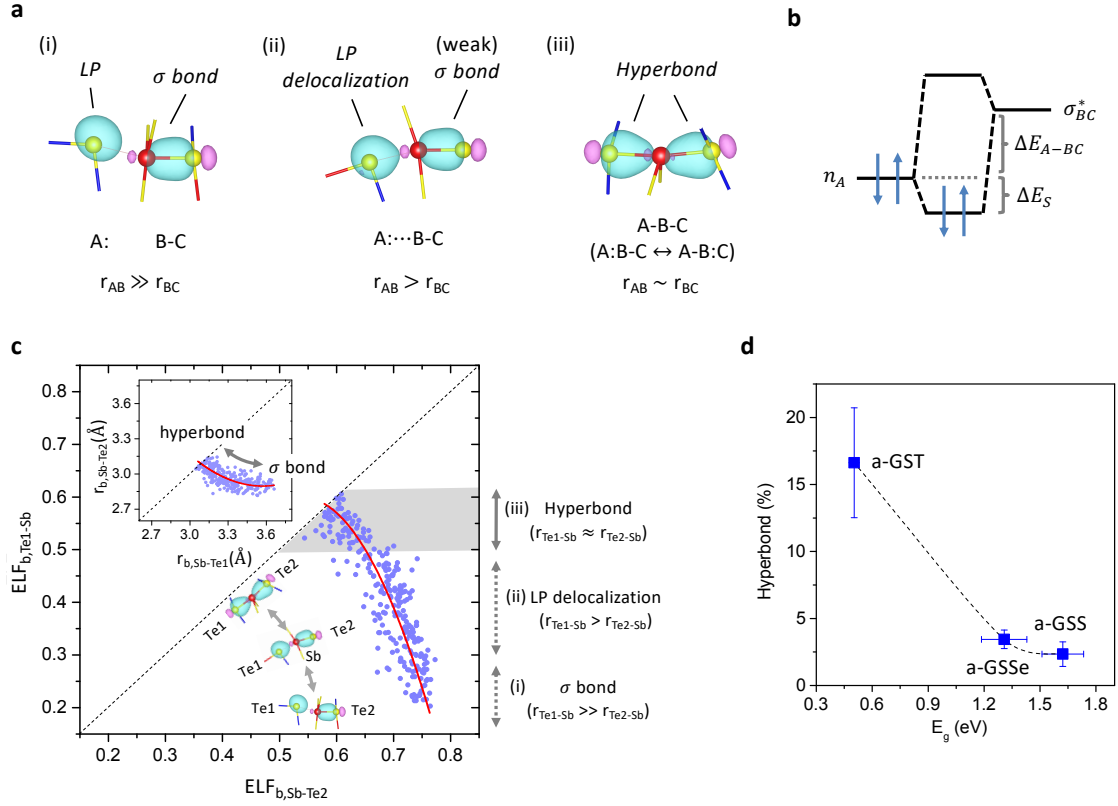

**Figure 2.** Mechanism of 3c/4e hyperbonding interactions in *a*-GST models. (a) 3c/4e bonding configurations for two limiting cases, (i) and (iii), and an intermediate interaction, (ii). (i) and (iii) correspond to an insignificant or a strong 3c/4e interaction, resulting in a covalent σ-bond or hyperbond pair formation, respectively. The LP ($n_A$) and bonding ($\sigma_{BC}$) orbitals are represented by isosurface plots of the maximally localized Wannier functions with light blue (positive) and pink (negative) contours. The antibonding orbital ($\sigma_{BC}^*$) is concentrated outside of the bonding region, thereby overlapping more with the LP than does $\sigma_{BC}$. The interaction (ii) reveals a non-trivial interaction between a LP orbital of A and the antibonding orbital of a B-C bonding pair. A: denotes a LP on an A atom, and ⋯ indicates a weak LP delocalization, while A-B-C denotes the formation of hyperbond pairs among A, B, and C atoms, which can be represented by the two ionic resonant structures, A:B-C and A-B:C. (b) A schematic interaction-energy diagram for a LP ($n_A$) stabilization interaction with a nearby antibonding orbital ($\sigma_{BC}^*$). (c) Variation of ELF values for both short- and long-distance interatomic interactions for (near) linear-bonding configurations found in *a*-GST models. The inset shows data points for interatomic distances rather than ELF values. The data points in the shaded area correspond to the hyperbonds defined in this study. (d) The ratio of 3c/4e hyperbonds to the ordinary 2c/2e covalent bonds for different chalcogen-containing amorphous models, depicted as a function of their band gaps. Here, the band gap was defined as the energy difference between the highest occupied and lowest unoccupied Kohn-Sham orbitals.

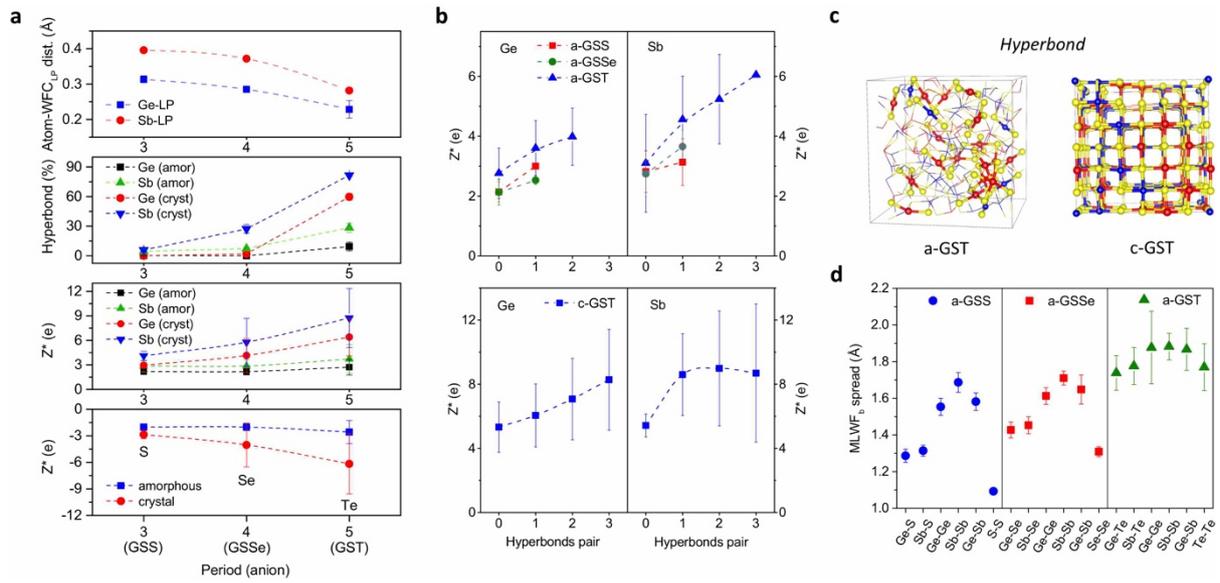

**Figure 3.** Comparison of amorphous and crystalline chalcogenide models. (a) Chalcogen-dependent property contrasts, showing the trend of varying *sp* hybridization, hyperbonding, and Born effective-charge (BEC) values. The extent of *sp* hybridization is represented by the averaged distance between the central atom (Ge or Sb) and its LP (top), and the percentage of hyperbonds (second from top) denotes the ratio of 3c/4e hyperbonds to ordinary 2c/2e covalent bonds. (b) Atomic species-resolved BECs, depending on the involved number of hyperbond pairs for amorphous (top) and crystalline (bottom) chalcogenide models. (c) Hyperbonds found in *a*-GST and *c*-GST with Ge (blue), Sb (red), and Te (yellow). (d) The spread of MLWFs for different types of bonding pairs for amorphous chalcogenide models.

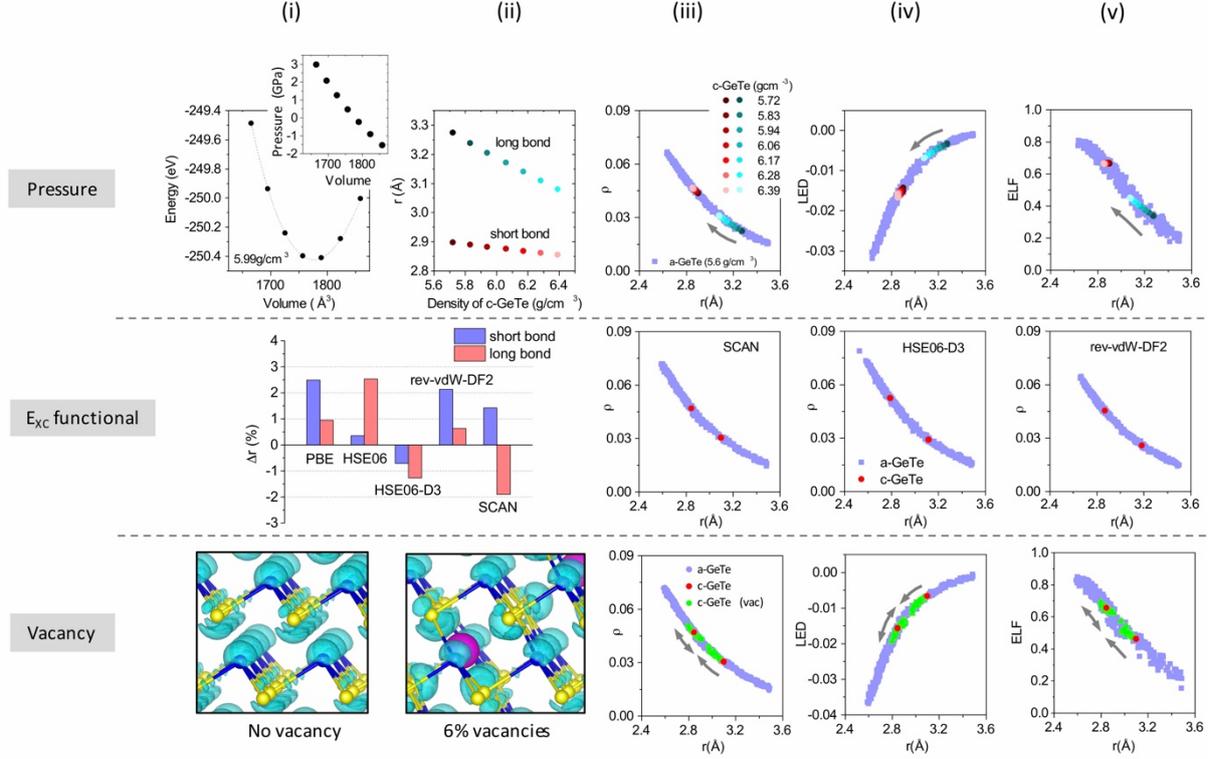

**Figure 4.** Chemical bonding in amorphous and crystalline GeTe models. (top row) Effect of pressure (or model density) on: (i) the model energy ; (ii) short and long interatomic distances; (iii) chemical-bonding indicators of charge density; (iv) energy density; and (v) ELF. (Middle row) Effect of different exchange-correlation functionals. Short and long interatomic distances are shown with reference to the experimental values, taken from Ref. [46] (i-ii). The charge densities at bond-critical points are shown for amorphous and crystalline GeTe models generated with: (iii) SCAN[47]; (iv) HSE06-D3[48]; and (v) rev-vdW-DF2[49] exchange-correlation functionals. (bottom row) Effect of cation vacancies on ELF distributions, without (i) or with (ii) vacancies, and on the chemical-bonding indicators of charge density (iii), energy density (iv), and ELF (v). Ge (blue) and Te (yellow) atoms, with spheres representing vacant sites (magenta), are displayed. The value of the ELF isosurface (cyan) is 0.5. The units of chemical indicators here are the same as in Figure 1.

Supporting Information

**Chemical bonding in chalcogenides: the concept of multi-centre hyperbonding**

T. H. Lee* and S. R. Elliott*

Department of Chemistry, University of Cambridge, Lensfield Road, Cambridge, CB2 1EW, UK

# 1. Methods

1) *Ab initio* molecular-dynamics (AIMD) simulations

We employed constant-volume *ab initio* simulations based on density-functional theory (DFT) with the generalized-gradient approximation (GGA), as implemented in the VASP code [1]. The projector augmented-wave (PAW) method [2] with the Perdew-Burke-Ernzerhof (PBE) [3] or the SCAN exchange-correlation functional [4] were used for the GST models. The SCAN [4], the HSE06-D3 [5], and the rev-vdW-DF2 [6] functionals were used for the GeTe models. The HSE06 hybrid functional [7] was used to calculate the band gap of the models. The outer s and p electrons of constituent atoms were treated as valence electrons. The plane-wave energy cutoff for the calculation of band gaps with the HSE06 functional was 400 eV, while the plane-wave energy cutoff was 300 eV for all other calculations. The Brillouin zone was sampled at the Γ point for amorphous models, while a Monkhorst-Pack 2x2x2 grid was used for crystalline models. For AIMD simulations, the temperature was controlled by a Nosé thermostat algorithm, and the MD time step was 3 fs. The amorphous and crystalline models were simulated in cubic supercells with periodic boundary conditions. Structural relaxation was performed using a conjugate-gradient method, until the force on any atom was below 0.01 eV/Å. The Born effective charges (BECs) were determined using density-functional perturbation theory, as implemented in the VASP code [1].

2) Amorphous and crystalline models

Amorphous models (315 atoms) of Ge-Sb-S (GSS), Ge-Sb-Se (GSSe) and Ge-Sb-Te (GST), all with the 225 composition, were generated via the conventional 'melt-quench' method. Initial random configurations of atoms were mixed at 2000 K for a few ps, and then kept at liquid temperature for tens of ps. For each type of chalcogenide, three amorphous models were generated by quenching different liquid configurations, sampled at different MD trajectories, with at least a 10 ps interval to ensure different initial quenching configurations. The quenching rate was -15 K/ps. The three different amorphous models were then used to collect statistics on static or dynamic material properties for each chalcogenide system. The density used for *a*-GST models was the experimental density of 5.88 g/cm$^3$. The densities of *a*-GSS and *a*-GSSe were determined via two steps. First, an amorphous model for each type of chalcogenide was generated with an approximate density, estimated with reference to the density of *a*-GST, and then the cell volume was relaxed to find the minimum-energy density. The determined densities of *a*-GSS and *a*-GSSe models were 3.43 and 4.41 g/cm$^3$, respectively. This density was then used to regenerate amorphous models by repeating the first step. The final pressure of models generated in this way was close to zero. Similarly, for statistics, three rocksalt-type crystalline models (each 196 atoms) for each crystalline chalcogenide were generated. The anion sites were filled with the relevant chalcogen atoms. The cation sites were randomly filled with Ge, Sb, and 20% of vacancies. For a proper comparison, three random distributions were used in common as initial configurations of c-GSS, c-GSSe, and c-GST models with approximate densities. Those three initial crystalline structures correspond to the perfect rocksalt-like structures without any distortion. However, structural (including a model volume) relaxation towards a minimum-energy configuration leads to a distortion of the local coordination, with the extent of the distortion strongly depending on the chalcogen type (most significant for GSS), as noted in the main text. These relaxed models were then analyzed to get an insight into the impact of *sp* hybridization of AOs on the hyperbonding tendency and on the crystal structures.

3) Chemical-bonding indicators and definition of bonds

Five different chemical-bonding indicators were analysed to characterize interatomic interactions. The electron charge density, its Laplacian, local energy density, and electron-localization function ($ELF_b$) were analysed at bond-critical points (BCPs) for the relevant atom pairs within the framework of Quantum Theory of Atoms in Molecules (QTAIM) [8]. The BCP for each pair of atoms was determined by the critic2 code [9]. The orbital-based indicator of ICOHP was determined by the LOBSTER code [10]. Maximally-localized Wannier functions and their spread were computed by the wannier90 code [11]. We analysed ELF attractors [12], representing LPs, which were then used to characterize the magnitude of *sp* hybridization of LPs for the crystalline models. We employed several different exchange-correlation functionals, namely PBE [3] and SCAN [4] for GST and GeTe models, and additionally HSE06-D3 [5] and rev-DFT-D2 [6] for GeTe models. Except for PBE, all the other functionals include the effect of van der Waals interactions. The same conclusions were reached for all functionals considered.

      The term 'hyperbond' was used specifically to represent each bond of a bond pair constituting a (near-)linear triatomic bonding geometry with near-identical bond distances, while the term '3c/4e bond' was used in a more relaxed manner, as related to the bond angles and distances. In this study, we define a hyperbond when $ELF_b$ for both linear bonds exceeds a value of 0.5 [13]. It should be emphasized, though, that other chemical-bonding indicators, e.g. charge density or ICOHP, can be equivalently employed, due to the definite relationships among those indicators (Figure 1). Similarly, although an ordinary covalent bond was defined when the corresponding $ELF_b$ value exceeds 0.5, other chemical-bonding indicators can be also used. In this study, we focused on Ge-Te and Sb-Te 'correct' bonds, rather than 'wrong'

(homopolar and Ge-Sb) bonds that are not observed in ideal crystals. As their concentrations, for linear triatomic bonding geometries, are relatively very low, they were not considered further here.

The bonding of Ge and Te atoms in rhombohedral *c*-GeTe consists of three strong (short) and three weak (long) Ge-Te bonds. Strictly speaking, these long (weak) bonds in *c*-GeTe do not satisfy the bonding criteria adopted in this study (Figure 4), but, for simplicity, they are called 'long bonds' in the main text. Models, generated with various exchange-correlation functionals, reproduced the experimental short and long bond lengths to within an error of $\sim\pm2\%$. The rhombohedral symmetry persists with pressure (Figure 4), but the length ratio of short to long bonds continuously diminishes. In particular, the reduction in length of long bonds is more pronounced than for short bonds.

## 2. Comparison between *a*- and *c*-GST models: Bader charge

In the main text, the interatomic interactions were characterized in terms of various chemical-bonding indicators [8-12] in order to emphasize the similarity of chemical-bonding interactions between amorphous and crystalline GST (and GeTe) models for the *same* bond distances. Also, from the intrinsic properties of hyperbonds (see the main text), the contrast in various microscopic material properties between amorphous and crystalline phases was elucidated: one of these properties is the difference in Bader charges. The formation of hyperbonds from the interaction between a LP orbital and an antibonding state of a nearby σ bond (Figure 2a) leads to an increase in the polarity of bonds (i.e. ionicity), thereby resulting in an enhanced charge transfer from Ge and Sb atoms to Te [13]. This charge transfer is manifested in higher Bader charges for *c*-GST accompanying the very significant increase in the number of hyperbonds. Specifically, the averaged Bader charges for *a*-GST (*c*-GST) were found to be 0.33 (0.37), 0.42 (0.48), and -0.30 (-0.35) for Ge, Sb, and Te, respectively. The distribution of Bader charges for each atomic species is also shown in Figure S1. Similar to the case of the chemical-bonding indicators, the range of Bader charges for *c*-GST is not completely separated from the distribution for *a*-GST, but both distributions are rather overlapped, since some of the atoms in *a*-GST are also involved in hyperbonding. Consequently, the distributions for *c*-GST are shifted to larger values of Bader charge in comparison with the distribution for *a*-GST, eventually resulting in the difference in the averaged Bader charges between *a*- and *c*-GST models. If atoms that are involved in hyperbonding are differentiated from the other atoms in *a*-GST, then the Bader charges for the former are found to be very similar to those found in *c*-GST, i.e. 0.37 (0.47) for hyperbonding Ge (Sb) atoms in *a*-GST. This result indeed indicates the different ionicity between normal covalent bonding and hyperbonding, as well as the similar nature of hyperbonds found in *a*- and *c*-GST.

**3. Chalcogen-dependent crystalline models**

1) Coordination change with relaxation

As noted in the main text, the degree of *sp* hybridization is a useful quantity for describing the structure of crystalline chalcogenides. For this, we compared structural-relaxation behaviour between crystalline models of different chalcogenides with the same initial rocksalt-like configurations. The local coordination change involved with the relaxation is mostly a transition from an initially octahedral coordination to the trigonal-pyramidal, seesaw, or square-pyramidal geometries, with the formation of a LP on each cation atom [13]. These transitions can be described by the sequential decomposition processes of (iii) → (ii) → (i) in Figure 2a, along three, two, or a single hyperbonding direction(s), respectively. The extent of local structural changes, inducing structural disorder, differs, depending on the models, in the order of increasing disorder of GST, GSSe, and GSS. For instance, 100% of Ge and 91% of Sb atoms in *c*-GSS transformed into trigonal pyramidal units, and 97% of Ge and 59% of Sb atoms did so in *c*-GSSe: this compares with only 20% of Ge and 2% of Sb doing the same in *c*-GST.

2) Interatomic-distance distributions

The distribution of interatomic distances for Ge-X and Sb-X bonds in *c*-GSX (where X = S, Se, or Te) are shown in Figure S2. It is clear from the figures that the differences between the peak positions (average distances) at short, and long, interatomic distances get smaller, as the sulphur chalcogen of GSS is replaced by Se and then by Te. Given that each of the two peak positions is considered as representing either a short or long bond length, the ratio of the long to short bond lengths can measure the magnitude of the structural distortion after structural relaxation. The estimated bond-length ratios for Ge-S (Sb-S), Ge-Se (Sb-Se), and Ge-Te (Sb-Te) are ~1.3 (1.2), ~1.1 (1.2), and ~1 (1), respectively, showing that the structural distortion becomes weaker, in the order GSS → GSSe → GST. This result can be interpreted as a consequence of the difference in *sp* hybridization of atomic orbitals involved in bonding (or non-bonding), as noted in the main text. The difference in hyperbonding tendency depending on the type of chalcogens equally explains this observation (see the main text).

3) ELF attractors for LPs

The *sp* hybridization of AOs in *c*-GST was investigated by analyzing the distance from the central atom to the ELF attractors representing LPs (Figure S3) [12]. The trend is similar to the case of *a*-GST, in which the position of LP MLWFs was instead used to measure the degree of *sp* hybridization, although the difference between *c*-GSSe and *c*-GST is not as significant as in the case of *a*-GST. This may be due to the fact that attractor positions are not sensitive enough to represent the position of LP centres, compared to MLWFs. Nevertheless, the ELF values at the attractors follow the trends expected with increasing metallicity, i.e. lower localization on proceeding down the column of the periodic table.

4) Bond-angle distributions

The bond-angle distributions (BADs) for different central atoms in different crystalline chalcogenide models are shown in Figure S4. The BAD peak positions for *c*-GSS and *c*-GSSe are both at ~95°, while those for *c*-GST are at ~92° for Ge, and at ~90° for Sb, although *c*-GST shows much broader distributions compared to *c*-GSS or *c*-GSSe. The larger width of the BAD for *c*-GST is due to the formation of linear triatomic bonding geometries, while the percentage of such bonding geometries is much smaller for *c*-GSS, and less so for *c*-GSSe. The narrower BADs for *c*-GSS and *c*-GSSe indicate that the bonding is stiffer than in *c*-GST.

**4. The mechanism of 3c/4e hyperbonding**

The mechanism of hyperbonding is described in Figure 2. Figure S5 supports the proposed hyperbonding mechanism by showing that the ligand Te atoms are mostly of the Te(3,1) type [13], that is, three-fold coordinated Te atoms with a single LP. The additional LP, compared to two-fold coordinated Te atoms, is involved in 3c/4e hyperbonding interactions, as illustrated in the main text.

**5. The spread of bonding MLWFs**

The spreads of bond MLWFs for different amorphous chalcogenide models are shown in Figure 3d. The larger standard deviations for *a*-GST than for *a*-GSS or *a*-GSSe models are attributed in the main text to the stronger delocalization of electrons involved in 3c/4e hyperbonding than that of electrons involved in ordinary 2c/2e covalent bonding, which is supported by the distribution of the spreads of bond MLWFs, as shown in Figure S6.

**6. Comparison with the valence-alternation pair (VAP) model**

The hyperbonding interaction involves a two-fold coordinated chalcogen atom (i.e. a normally bonded chalcogen) and a (bonded) pair of atoms. However, the VAP model[14] involves a normally two-fold coordinated chalcogen atom, and a positively charged, three-fold coordinated chalcogen-atom defect, and a negatively charged, one-fold coordinated chalcogen-atom defect. For both interactions, one LP from the two-fold coordinated chalcogen is involved in forming a bond, but their interacting orbital, and the final products, are completely different. In the case of the hyperbonding interaction, the LP interacts with the antibonding orbital of a neighbouring atom pair through a two-electron stabilization interaction, in which the LP and antibonding orbitals are interacting over the three involved atoms. Therefore, this interaction is a three-centre, four-electron (3c/4e)-type of bonding. As an outcome of this interaction, either two equal bonds (Fig. 2a (iii)), or a pair of short and long bonds (Fig. 2a (ii)) with a (near) linear bonding geometry, are formed. On the other hand, in the case of the valence-alternation pair (VAP), the LP interacts with the empty *p* orbital of a positively charged, one-fold coordinated chalcogen atom, forming an ordinary two-centre, two-electron (2/2e) covalent bond (or, more precisely, coordinate bond) through the strong electron-phonon coupling. The outcome of this interaction is therefore a pair of defects, a negatively charged, one-fold coordinated chalcogen atom ($C_1^-$), and a positively charged, three-fold coordinated chalcogen atom ($C_3^+$), i.e. the VAP, and normally-bonded, neutral, two-fold coordinated chalcogen atoms ($C_2^0$). A variety of interesting properties of hyperbonds originate from the multi-centre nature of interactions, and also, in part, from their unique, linear-bonding geometries.

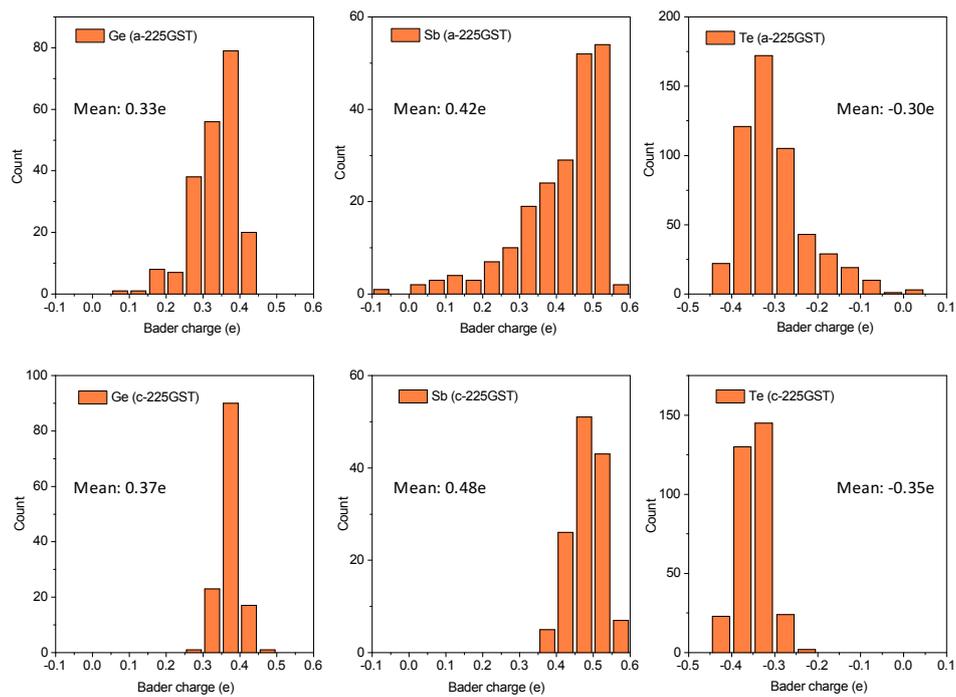

Figure S1. Comparison of Bader-charge distributions for constituent atoms in *a*- and *c*-GST.

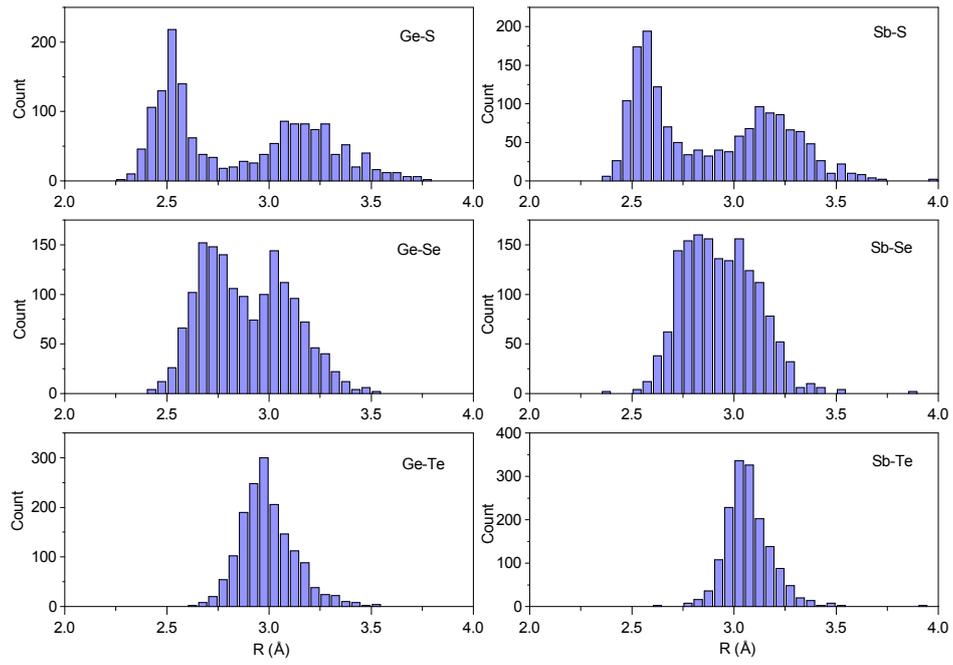

Figure S2. Interatomic-distance distributions for different crystalline chalcogenide models: *c*-GSS (top), *c*-GSSe (middle), and *c*-GST (bottom).

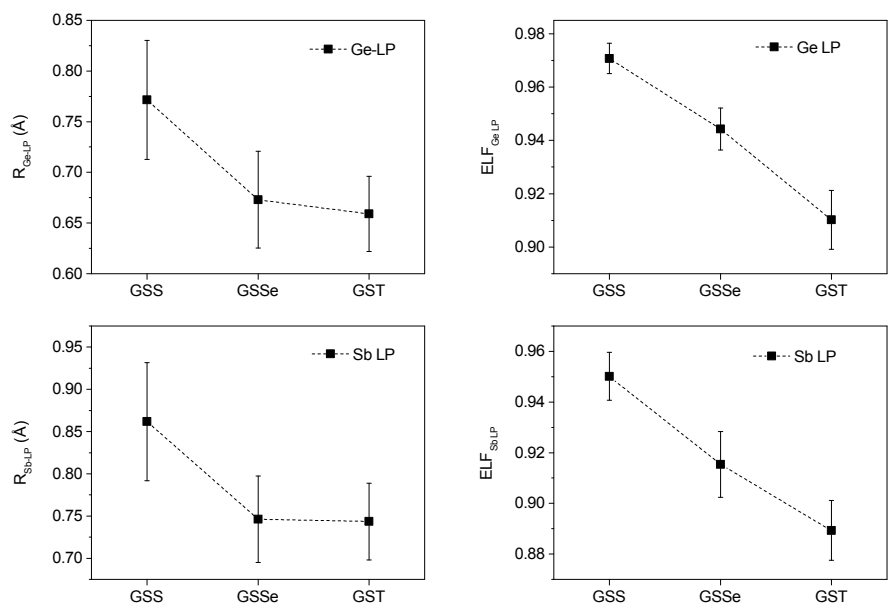

Figure S3. Atom-LP distances (left) and ELF values (right) at LP attractors in crystalline models of GSS, GSSe and GST.

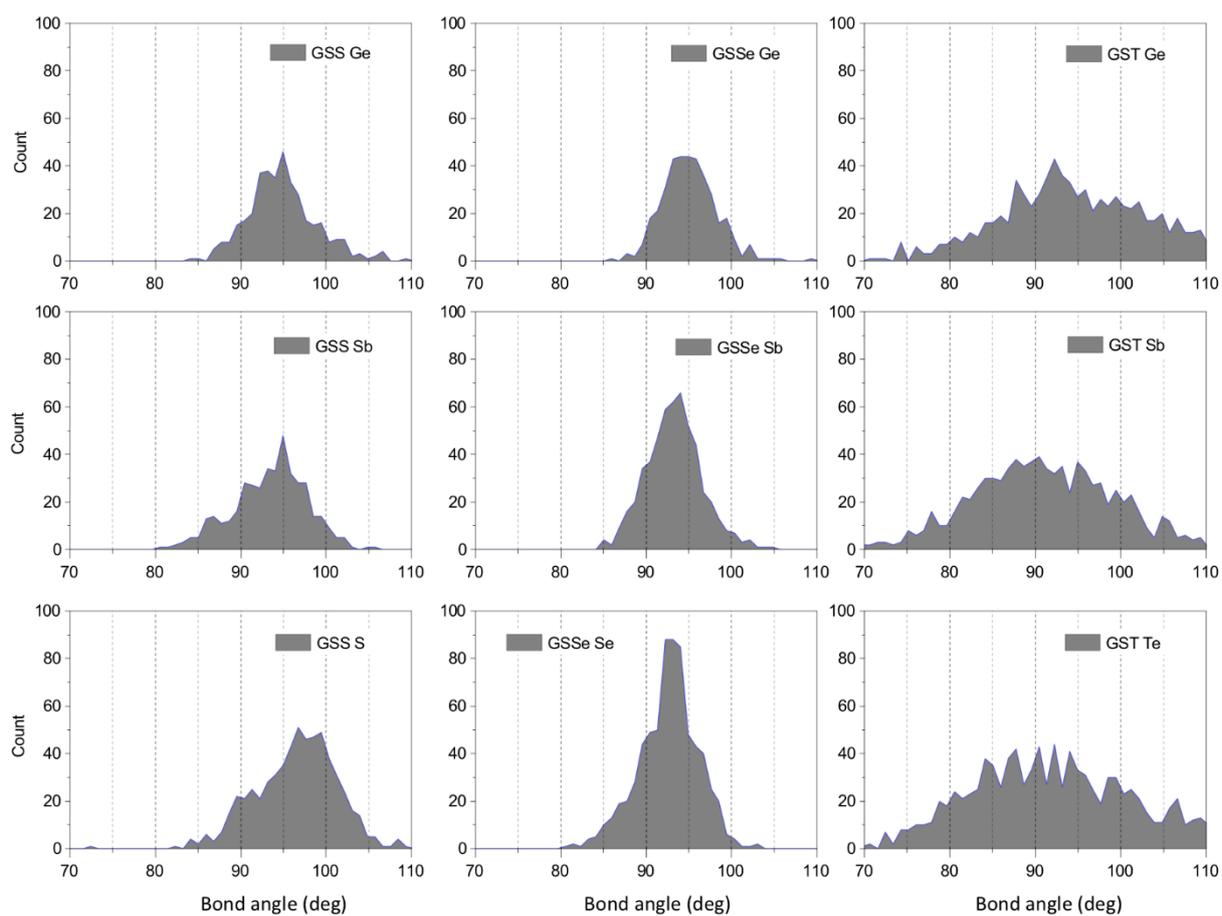

Figure S4. Bond-angle distributions around constituent atoms for different crystalline chalcogenide models of GSS, GSSe and GST.

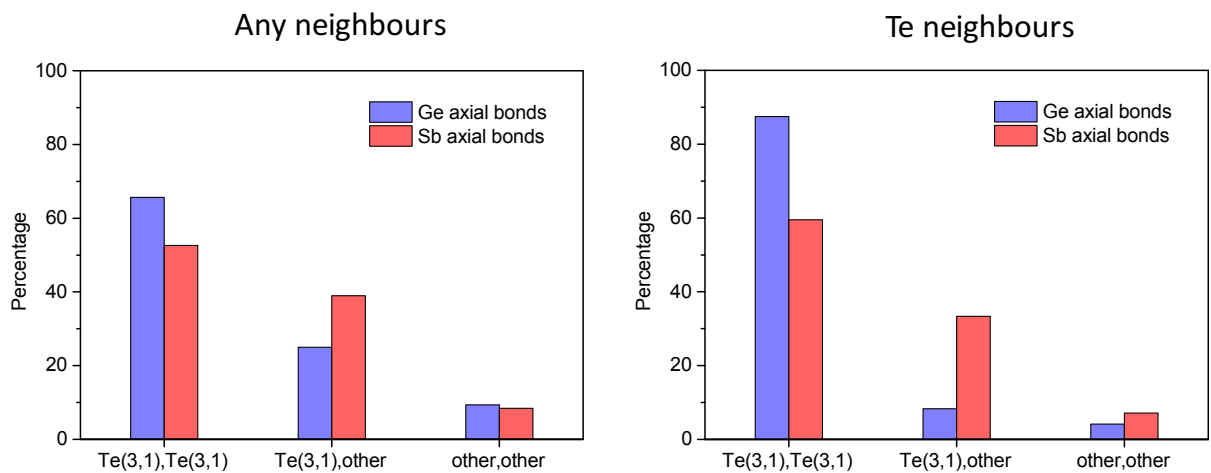

Figure S5. Types of ligand atoms for axial bonds in *a*-GST models. Ge (blue) or Sb (red) atoms residing at the centre of a pair of linear axial bonds are coordinated by two ligand atoms. For each ligand atom, there exist various forms of structural motifs (units) [13]. In particular, we focused only on Te(3,1)-type units among others, as this unit can provide LP electrons for the formation of hyperbonds (see the main text). For both panels above, the first data (denoted as 'Te(3,1),Te(3,1)') indicate that both ligand atoms are Te(3,1)-type units. Similarly, the second data (i.e. 'Te(3,1), other') correspond to the case where one of the two ligand atoms is a Te(3,1), and the other ligand corresponds to a structural unit other than Te(3,1), while the last data (i.e. 'other, other') indicate the case where both ligands are structural units other than Te(3,1) units. All possible types of structural units were considered as 'other' ligand types for the left panel. On the other hand, for the right panel, the percentages were calculated among axial bonds consisting only of Te ligand atoms.

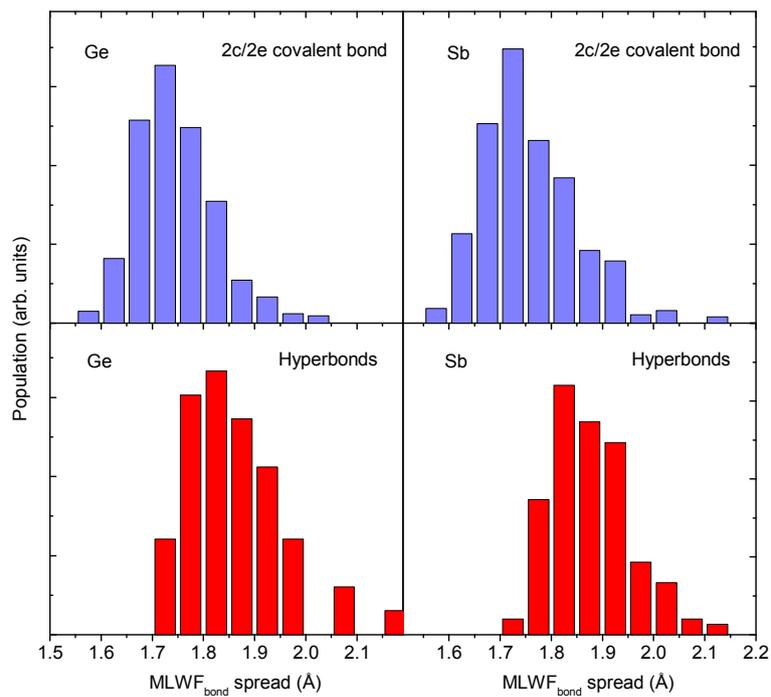

Figure S6. Spread of MLWF values for ordinary covalent (2c/2e) bonds and hyperbonds in *a*-GST models.

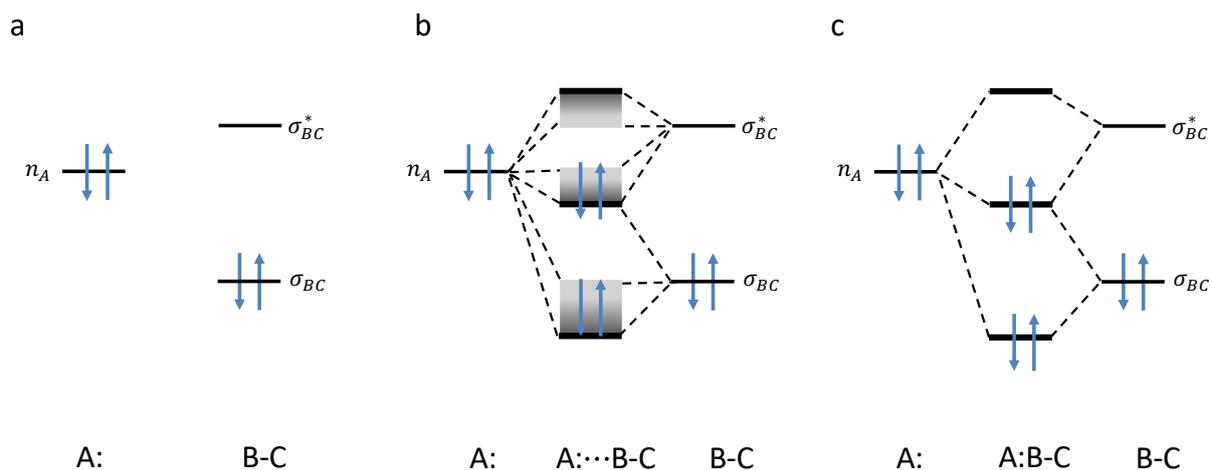

Figure S7. Schematic energy-level diagrams for 3c/4e hyperbonding interactions. Energy-level diagram for the configuration: a) (i) in Fig. 2a; (b) for (ii); and (c) for (iii). A: denotes a LP on an A atom, and ⋯ indicates a weak LP delocalization. The formation of a pair of hyperbonds among A, B, and C atoms can be represented by the two ionic resonant structures, A:B-C and A-B:C. The energy bands in (b) indicates different strengths of hyperbonding interactions with varying extents of LP delocalization.

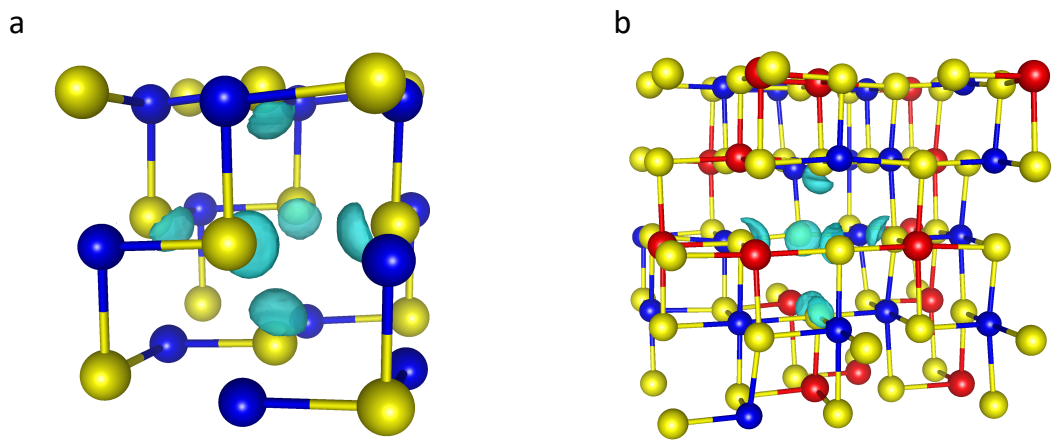

Figure S8. Plots of electron-localization function (ELF) isosurfaces near vacancies. After a vacancy was generated at the centre of six nearest-neighbour Te atoms in *c*-GeTe (a) and in *c*-GST (b), volumes enclosed by high-ELF isosurfaces (cyan) (ELF > 0.88) appear near these Te atoms. ELF isosurfaces only near the Te atoms (within a spherical radius of 1.5 Å from each Te atom) were plotted with Ge (blue), Sb (red), and Te (yellow).